\documentclass{article}%
\usepackage{amsfonts}
\usepackage{amsmath}%
\usepackage{amssymb}%
\usepackage{graphicx}
\newtheorem{theorem}{Theorem}

\newtheorem{definition}[theorem]{Definition}
\newtheorem{example}[theorem]{Example}

\newtheorem{lemma}[theorem]{Lemma}

\newtheorem{proposition}[theorem]{Proposition}
\newtheorem{remark}[theorem]{Remark}

\newenvironment{proof}[1][Proof]{\textbf{#1.} }{\ \rule{0.5em}{0.5em}}
\begin{document}

\title{Dirac reduction of dual Poisson-presymplectic pairs}
\author{Maciej B\l aszak\thanks{Partially supported by The Swedish Institute
scholarship No. 03824/2003}\\Institute of Physics, A. Mickiewicz University\\Umultowska 85, 61-614 Pozna\'{n}, Poland
\and Krzysztof Marciniak\thanks{Partially supported by The Swedish Research Council
grant No. 624-2003-607}\\Department of Science and Technology \\Campus Norrk\"{o}ping, Link\"{o}ping University\\601-74 Norrk\"{o}ping, Sweden}
\maketitle

\begin{abstract}
A new notion of a dual Poisson-presymplectic pair is introduced and it
properties are examined. The procedure of Dirac reduction of Poisson operators
onto submanifolds proposed by Dirac is in this paper embedded in a geometric
procedure of reduction of dual \ Poisson-presymplectic pairs. The method
presented generalizes those used by Marsden and Ratiu for reductions of
Poisson manifolds. Two examples are given.

\end{abstract}


\section{Introduction}

In \cite{Dirac} P.A.M. Dirac introduced a method of reducing a given Poisson
bracket onto a submanifold $\mathcal{S}$ given by some constraints $\varphi$.
A geometric meaning of this reduction procedure has been investigated in
\cite{MarsdenRatiu} and also in \cite{Diracrevisited}. In this paper we
complete this picture by its "dual" part by developing a theory of
Marsden-Ratiu type reduction of presymplectic 2-forms $\Omega$ that are (in a
sense developed below) dual to a given Poisson operator $\Pi$.

This paper is organized as follows. In this section we recall some basic
notions from Poisson and presymplectic geometry. In Section 2 we introduce and
discuss a central for this paper notion of \ a dual Poisson-presymplectic
(dual P-p) pair. We also examine some basic properties of P-p pairs. In
Section 3 we present a geometric reduction procedure of such a pairs to any
submanifold that its tangent bundle contains the kernel of the presymplectic
form that enters our P-p pair. This is the main section of this paper. We
conclude the article by Section 4 containing two examples.

Given a manifold $\mathcal{M}$, a \emph{Poisson operator} $\Pi$ on
$\mathcal{M}$ is a bivector, $\Pi\in\Lambda^{2}(\mathcal{M})$ (degenerate in
general) such that its Schouten bracket with itself vanishes. A function
$c:\mathcal{M}\rightarrow\mathbf{R}$ is called \emph{Casimir function} of the
Poisson operator $\Pi$ if for any function $F:\mathcal{M}\rightarrow
\mathbf{R}$ we have $\left\{  F,c\right\}  _{\Pi}=0$ (or, equivalently, if
$\Pi dc=0$). A vector field $X_{f}$ related to a function $f$ by relation
\begin{equation}
X_{f}=\Pi df\label{1.1}%
\end{equation}
is called a Hamiltonian vector field with respect to the Poisson operator
$\Pi$. If $X$ is any vector field on $\mathcal{M}$ that is Hamiltonian with
respect to $\Pi$ then $L_{X}\Pi=0$, where $L_{X}$ is the Lie-derivative
operator in the direction $X$.

Further, a \emph{presymplectic form} $\Omega$ on $\mathcal{M}$ is a 2-form
that is closed (degenerate in general). The kernel of any presymplectic form
of constant rank is always integrable. A vector field $X_{f}$ related to a
function $f$ by relation%
\begin{equation}
\Omega X_{f}=df\label{1.2}%
\end{equation}
is called an inverse Hamiltonian vector field with respect to the
presymplectic operator $\Omega$. For a closed two-form $\Omega$ if
$\Omega(Y)=0$ for some vector field $Y$ on $\mathcal{M}$ then $L_{Y}\Omega=0$.

Notice that when $\Pi$ is nondegenerate one can always define $\Omega=\Pi
^{-1}$, and then the equations (\ref{1.1}) and (\ref{1.2}) are equivalent and
a vector field that is Hamiltonian with respect to $\Pi$ is simultaneously
inverse Hamiltonian with respect to $\Omega.$ In the degenerate case we
encounter problems. Firstly, one can not define $\Omega$ as the inverse of
$\Pi.$ Secondly, for a degenerate $\Pi$ the equation (\ref{1.1}) defines a
Hamiltonian vector field for any function $f$ (as in the nondegenerate case),
while for a degenerate $\Omega$ and arbitrary $f$ there is usually no such
vector field $X_{f}$ that (\ref{1.2}) is fulfilled. In other words, equation
(\ref{1.2}) is valid only for a particular class of functions (contrary to the
nondegenerate case). We will try to overcome these difficulties in the next
Section. We will constantly assume that our degenerate operators are of
constant rank.

\section{Dual Poisson-presymplectic pairs}

In this section we develop and discuss a central for this article notion of a
dual Poisson-presymplectic pair.

Consider a smooth manifold $\mathcal{M}$ of dimension $m$ equipped with a pair
of antisymmetric operators $\Pi$, $\Omega$.

\begin{definition}
\label{dualdef}A pair of antisymmetric tensor fields $(\Pi,\Omega)$ such that
$\Pi:T^{\ast}\mathcal{M}\rightarrow T\mathcal{M}$ (i.e. $\Pi$ is twice
contravariant) and $\Omega:T\mathcal{M}\rightarrow T^{\ast}\mathcal{M}$ (i.e.
$\Omega$ is twice covariant) is called a dual pair if there exists $r$, $0\leq
r\leq m$, functionally independent scalar functions $c_{i}:\mathcal{M}%
\rightarrow\mathbf{R}$, $i=1,\ldots,r$ and $r$ linearly independent vector
fields $Y_{i}$, $i=1,\ldots,r$ such that the following conditions are satisfied:

\begin{enumerate}
\item $Y_{i}(c_{j})=\delta_{ij}$ for all $i,j=1,\ldots r$.

\item The kernel of $\Pi$ is spanned by the differentials $dc_{i}$, $\ker
(\Pi)=Sp\{dc_{i}\}_{i=1..r}$.

\item The kernel of $\Omega$ is spanned by the vector fields $Y_{i}$,
$\ker(\Omega)=Sp\{Y_{i}\}_{i=1..r}$.

\item The following\emph{\ partition of identity} holds on $T\mathcal{M}$%
\begin{equation}
I=\Pi\Omega+%
{\textstyle\sum\limits_{i=1}^{r}}
Y_{i}\otimes dc_{i}\label{partition}%
\end{equation}
where $\otimes$ denotes the tensor product.
\end{enumerate}
\end{definition}

We would like to point out that our definition of a dual pair means something
different than the definition of a dual pair introduced by A. Weinstein in
\cite{weinstein}. Besides, we mostly work with dual pairs consisting of
Poisson operators and closed forms (which we call 'dual P-p pairs', see
below), so hopefully it will not cause any confusion.

The partition of identity (\ref{partition}) reads on $T^{\ast}\mathcal{M}$ as%
\[
I=\Omega^{\ast}\Pi^{\ast}+%
{\textstyle\sum\limits_{i=1}^{r}}
dc_{i}\otimes Y_{i}%
\]
which due to antisymmetry of $\Pi$ and $\Omega$ yields
\begin{equation}
I=\Omega\Pi+%
{\textstyle\sum\limits_{i=1}^{r}}
dc_{i}\otimes Y_{i}.\label{partitiondual}%
\end{equation}

\bigskip

Let us call the foliation of $\mathcal{M}$ associated with the functions
$c_{i}$ by $\mathcal{N}$. That foliation consists of level submanifolds
$\mathcal{N}_{\nu}$ of functions $c_{i}$ , $\mathcal{N}_{\nu}=\{x\in
M:c_{i}(x)=\nu_{i}$, $i=1,\ldots,r\}$, $\nu=(\nu_{r},\ldots,\nu_{r})$. The
condition 1 of the above definition implies that the distribution $Y$ spanned
by the vector fields $Y_{i}$ is transversal to the foliation $\mathcal{N}$
i.e. that no vector in $Y$ is ever tangent to the foliation $\mathcal{N}$.
Thus, for any $x\in\mathcal{M}$ we have%
\[
T_{x}\mathcal{M}=T_{x}\mathcal{N}_{\nu}\oplus Y_{x}\text{, \ \ }T_{x}^{\ast
}\mathcal{M}=T_{x}^{\ast}\mathcal{N}_{\nu}\oplus Y_{x}^{\ast}%
\]
where $\mathcal{N}_{\nu}$ is a submanifold from the foliation $\mathcal{N}$
that passes through $x$, the symbol $\oplus$ denotes the direct sum of the
vector spaces, $Y_{x}$ is the subspace of $T_{x}\mathcal{M}$ spanned by the
vectors $Y_{i}$ at this point, $T_{x}^{\ast}\mathcal{N}_{\nu}$ is the
annihilator of $Y_{x}$ and $Y_{x}^{\ast}$ is the annihilator of $T_{x}%
\mathcal{N}_{\nu}$. The condition 2 of the above definition implies that the
image $\operatorname{Im}(\Pi)$ is at every point tangent to a level
submanifold $\mathcal{N}_{\nu}$ that passes through this point. Indeed, if
$\Pi dc_{i}=0 $ then for any 1-form $\alpha$ we have, due to the antisymmetry
of $\Pi$, that $\left\langle dc_{i},\Pi\alpha\right\rangle =-\left\langle
\alpha,\Pi dc_{i}\right\rangle =0$ for all $i=1,\ldots,r$, so that the vector
$\Pi\alpha$ is always tangent to $\mathcal{N}$. The condition 3 means that
$\operatorname{Im}(\Omega)$ is in every point $x$ contained in $T_{x}^{\ast
}\mathcal{N}_{\nu}$ (again for appropriate $\nu$). Indeed, if $\Omega
(Y_{i})=0$ for all $i=1,\ldots,r$ then for any vector field $V$ we have (due
to the antisymmetry of $\Omega$) $\ \left\langle \Omega V,Y_{i}\right\rangle
=-\ \left\langle \Omega Y_{i},V\right\rangle =0$. The condition 4 is the most
interesting one: obviously, it describes the degree of degeneracy of our pair.
But if we restrict our attention to those dual pairs that consist of a Poisson
operator and of a closed 2-form then it has yet another, deeper meaning.

\begin{definition}
\label{Ppdef}A dual pair $(\Pi,\Omega)$ is called a dual Poisson-presymplectic
pair (in short: dual \emph{P-p }pair) if $\Pi$ is Poisson and if $\Omega$ is closed.
\end{definition}

\begin{remark}
In the case where a dual P-p pair has no degeneration ($r=0$) we get the usual
Poisson-symplectic pair of mutually inverse operators, since (\ref{partition})
reads then as $I=\Pi\Omega$. In the case where $r=m$ we have full
degeneration: $\Pi=0$ and $\Omega=0$ as then $Sp\{dc_{i}\}_{i=1..r}=T^{\ast
}\mathcal{M}$ and $Sp\{Y_{i}\}_{i=1..r}=T\mathcal{M}$. This case will be
therefore excluded as non-interesting.
\end{remark}

\bigskip Let $(\Pi,\Omega)$ be a dual P-p pair and let%
\begin{equation}
X_{f}=\Pi df\label{2.1}%
\end{equation}
be a Hamiltonian vector field with respect to $\Pi.$ Applying $\Omega$ to both
sides of (\ref{2.1}) we get%
\begin{equation}
df=\Omega(X_{f})+%
{\displaystyle\sum\limits_{i=1}^{r}}
Y_{i}(f)dc_{i}.\label{2.2}%
\end{equation}
In that sense $\Omega$ plays the role of the "inverse" of $\Pi.$ Notice that
vector fields that are Hamiltonian with respect to $\Omega$ are precisely
those that are related to functions $f$ which are annihilated by $\ker
(\Omega).$ For such functions (\ref{2.2}) reduces to (\ref{1.2}).

\begin{proposition}
For a dual P-p pair $(\Pi,\Omega)$ the vector fields $Y_{i}$ mutually commute:
$[Y_{i},Y_{j}]=0,\ i,j=1,\ldots,r$.
\end{proposition}

\begin{proof}
Since $\Omega$ is presymplectic, $\ker(\Omega)$ is an integrable distribution
so that
\[
\lbrack Y_{i},Y_{j}]=\sum_{k=1}^{r}\phi_{ij}^{k}Y_{k},
\]
where $\phi_{ij}^{k}$ are some functions on $\mathcal{M}$. Evaluating this
relation on all Casimirs $c_{l}$ we immediately find that $\phi_{ij}^{k}=0$
for all $i,j,k=1,\ldots,r$.
\end{proof}

Of course, the vector fields $Y_{i}$ and the forms $dc_{i}$ in the definition
of a dual pair are not unique. For example, we can change the basis in the
distribution spanned by $Y_{i}$ and compensate it by a change of basis in the
distribution spanned by $dc_{i\text{.}}$ We have however the following
uniqueness theorem:

\begin{theorem}
(uniqueness theorem) Suppose that $(\Pi,\Omega)$ and $(\Pi,\Omega^{\prime})$
are two dual P-p pairs that share the same $c_{i}$ and such that $\ker
(\Omega)=\ker(\Omega^{\prime})$. Then $\Omega=\Omega^{\prime}$.
\end{theorem}

\begin{proof}
Since $\ker(\Omega)=\ker(\Omega^{\prime})$ then $Y_{i}^{\prime}=%
{\textstyle\sum\limits_{j=1}^{r}}
\lambda_{ij}Y_{j}$ for some functions $\lambda_{ij} $ such that $\det
(\lambda_{ij})\neq0$. Thus%
\[
\delta_{ij}=Y_{i}^{\prime}(c_{j})=%
{\textstyle\sum\limits_{s=1}^{r}}
\lambda_{is}Y_{s}(c_{j})=%
{\textstyle\sum\limits_{s=1}^{r}}
\lambda_{is}\delta_{sj}=\lambda_{ij}\text{,}%
\]
so that $Y_{i}^{\prime}=Y_{i}$ for all $i$. Thus, since $I=\Pi\Omega+%
{\displaystyle\sum\limits_{i=1}^{r}}
Y_{i}\otimes dc_{i}$ and $I=\Pi\Omega^{\prime}+%
{\displaystyle\sum\limits_{i=1}^{r}}
Y_{i}\otimes dc_{i}$ we have that $\Pi\left(  \Omega^{\prime}-\Omega\right)
=0$ which implies $\Omega^{\prime}-\Omega=0$ since the product of two
antisymmetric operators is zero only if (at least) one of them is zero.
\end{proof}

The question thus arises: what is the actual 'gauge freedom' for a given dual
P-p pair? In other words: given a dual P-p pair $(\Pi,\Omega)$ how can we
deform $\Omega$ to a new presymplectic form $\Omega^{\prime}$ so that
$(\Pi,\Omega^{\prime})$ is again dual or how can we deform $\Pi$ to a new
Poisson operator $\Pi^{\prime}$ so that $(\Pi^{\prime},\Omega)$ is also a dual
P-p pair? An example of such a gauge freedom is given below.

\begin{proposition}
Let $(\Pi,\Omega)$ be a dual P-p pair as in definitions \ref{dualdef} and
\ref{Ppdef}. Define
\[
\Omega^{\prime}=\Omega+%
{\textstyle\sum\limits_{i}}
df_{i}\wedge dc_{i},
\]
where $f_{i}$ are some real functions on $\mathcal{M}$. Then $(\Pi
,\Omega^{\prime})$ is a dual P-p with $\ker(\Omega^{\prime})=Sp\left\{
Y_{i}^{\prime}=Y_{i}-\Pi\,df_{i}\right\}  $ provided that
\begin{equation}
Y_{i}(f_{j})-Y_{j}(f_{i})+\left\{  f_{i},f_{j}\right\}  _{\Pi}=0\text{ for all
}i,j\text{. }\label{gaugewar}%
\end{equation}

\end{proposition}

The proof is by direct computation.

Before we consider a gauge freedom for the operator $\Pi$ we prove a useful lemma.

\begin{lemma}
\label{l1 copy(1)}Let $(\Pi,\Omega)$ be a dual P-p pair. Then
\begin{equation}
L_{Y_{i}}\Pi=0,~\ \ \ \ \ i=1,\ldots,r.\label{property}%
\end{equation}

\end{lemma}

\begin{proof}
From the partition of identity and the property $L_{Y_{i}}\Omega=0$ we have
\begin{align*}
0  & =L_{Y_{i}}I=(L_{Y_{i}}\Pi)\Omega+\Pi(L_{Y_{i}}\Omega)+\sum_{j=1}%
^{r}[Y_{i},Y_{j}]\otimes dc_{j}\\
& =(L_{Y_{i}}\Pi)\Omega.
\end{align*}
On the other hand, from the property $\Pi dc_{j}=0,$ it follows that
\[
0=L_{Y_{i}}(\Pi dc_{j})=(L_{Y_{i}}\Pi)dc_{j}\ \ \ \text{as }\ \ L_{Y_{i}%
}dc_{j}=d(\delta_{ij})=0.
\]
Thus, from the decomposition (\ref{2.2}), for any function $f$ we have
\[
(L_{Y_{i}}\Pi)df=(L_{Y_{i}}\Pi)\left(  \Omega(X_{f})+%
{\displaystyle\sum\limits_{i=1}^{r}}
Y_{i}(f)dc_{i}\right)  =0
\]
and arbitrariness of $f~$implies that $L_{Y_{i}}\Pi=0$.
\end{proof}

\begin{proposition}
\label{guage}Suppose that $(\Pi,\Omega)$ is a dual P-p pair. Suppose that
$K_{i}$, $i=1,\ldots,r$ are vector fields that are Hamiltonian with respect to
$\Pi$ and inverse Hamiltonian with respect to $\Omega,$ i.e.
\begin{equation}
\Omega(K_{i})=dH_{i},\ \ \ \ K_{i}=\Pi dH_{i}\label{conditions}%
\end{equation}
for some functions $H_{i}$ and such that%
\begin{equation}
\Omega(K_{i},K_{j})=0\ \ \text{ for all }i,j.\label{gaugewar2}%
\end{equation}
Then the pair $(\Pi^{\prime},\Omega)$ with $\Pi^{\prime}=\Pi+%
{\textstyle\sum\limits_{i=1}^{r}}
Y_{i}\wedge K_{i}$ is a dual P-p pair with vector fields $Y_{i}$ and Casimirs
$c_{i}^{\prime}=c_{i}+H_{i}$.
\end{proposition}

\begin{proof}
An easy calculation with the use of partition of identity and the assumptions
(\ref{conditions}) yields
\[
\Pi^{\prime}\Omega=I-\sum_{i}Y_{i}\otimes dc_{i}^{\prime},
\]
so that the partition of identity for $(\Pi^{\prime},\Omega)$ is satisfied.
Moreover,%
\[
Y_{i}(H_{j})=\left\langle \Omega K_{j},Y_{i}\right\rangle =-\left\langle
\Omega Y_{i},K_{j}\right\rangle =0
\]
as $\Omega(Y_{i})=0$, so that
\[
Y_{i}(c_{j}^{\prime})=Y_{i}(c_{j})+Y_{i}(H_{j})=Y_{i}(c_{j})=\delta_{ij}.
\]
Further: $K_{i}(c_{j})$ $=\left\langle dc_{j},\Pi dH_{i}\right\rangle
=-\left\langle dH_{i},\Pi dc_{j}\right\rangle =0$ which implies
\[
\Pi^{\prime}dc_{j}^{\prime}=\sum_{i}\Omega(K_{j},K_{i})Y_{i}=0
\]
due to the assumption (\ref{gaugewar2}).

One can also show (by using . Lemma \ref{l1 copy(1)}) that the Schouten
bracket $[\Pi^{\prime},\Pi^{\prime}]_{S}$ vanishes so that $\Pi^{\prime}$ is
indeed Poisson.
\end{proof}

Let us now turn our attention to brackets induced on the space $C^{\infty
}(\mathcal{M})$ of all smooth real valued functions on $\mathcal{M}$.

We know that the Poisson operator $\Pi$ turns the space $C^{\infty
}(\mathcal{M})$ of all smooth real valued functions on $\mathcal{M}$ into a
Poisson algebra with the Poisson bracket (
\begin{equation}
\left\{  F,G\right\}  _{\Pi}=\left\langle dF,\Pi\,dG\right\rangle
\label{bracket}%
\end{equation}
In the case where $\Omega$ is a part of a dual P-p pair we can define the
above bracket through the action of $\Omega$ on $X_{F}$ and $X_{G}$.

\begin{proposition}
\label{rownenawiasy}Let $(\Pi,\Omega)$ be a dual P-p pair. Define a new
bracket on $C^{\infty}(\mathcal{M})$ $\ $through%
\[
\left\{  F,G\right\}  ^{\Omega}=\Omega(X_{F},X_{G}),
\]
where as usual $X_{F}=\Pi dF$ and $X_{G}=\Pi dG$. Then $\left\{  \cdot
,\cdot\right\}  ^{\Omega}=\left\{  \cdot,\cdot\right\}  _{\Pi}$ i.e. both
brackets are identical.
\end{proposition}

\begin{proof}
A simple calculation yields%
\begin{align*}
\Omega(X_{F},X_{G})  & =\left\langle \Omega X_{F},X_{G}\right\rangle
=\left\langle \Omega\Pi dF,\Pi dG\right\rangle \overset{\ast}{=}\\
& =\left\langle dF,\Pi dG\right\rangle -%
{\textstyle\sum\limits_{i}}
Y_{i}(F)\left\langle dc_{i},\Pi dG\right\rangle =\left\langle dF,\Pi
dG\right\rangle ,
\end{align*}
as $\left\langle dc_{i},\Pi dG\right\rangle =-\left\langle dG,\Pi
dc_{i}\right\rangle =0$. The equality with the asterisk is due to the
partition of identity on $T^{\ast}\mathcal{M}$ (\ref{partitiondual}).
\end{proof}

Let us now present two examples of P-p pairs.

\begin{example}
\label{pierwszy}Consider a manifold $\mathcal{M}$ parametrized locally by
coordinates
\[
(q_{1},\ldots,q_{n},p_{1},\ldots,p_{n},c_{1},\ldots,c_{r})
\]
and a pair of operators that in these coordinates have the form%
\begin{equation}
\Pi=\left[
\begin{array}
[c]{c|c}%
\begin{array}
[c]{cc}%
0_{n} & I_{n}\\
-I_{n} & 0_{n}%
\end{array}
& 0_{2n\times r}\\\hline
0_{r\times2n} & 0_{r}%
\end{array}
\right]  \text{ \ , \ }\Omega=\left[
\begin{array}
[c]{c|c}%
\begin{array}
[c]{cc}%
0_{n} & -I_{n}\\
I_{n} & 0_{n}%
\end{array}
& 0_{2n\times r}\\\hline
0_{r\times2n} & 0_{r}%
\end{array}
\right] \label{piomega1}%
\end{equation}
i.e. $\Pi$ is the canonical Poisson operator with $r$ Casimirs $c_{i}$ while
$\Omega$ is the canonical presymplectic form with the kernel spanned by
$Y_{i}=\frac{\partial}{\partial c_{i}}$. Here and in what follows $I_{n}$
denotes an $n\times n$ identity matrix and in general subscripts by a matrix
block denote the dimensions of this block. Obviously, $Y_{i}(c_{j}%
)=\delta_{ij}$. Moreover, the product $\Pi\Omega$ has the form%
\[
\Pi\Omega=\left[
\begin{array}
[c]{c|c}%
I_{2n} & 0_{2n\times r}\\\hline
0_{r\times2n} & 0_{r}%
\end{array}
\right]
\]
while the tensor product $%
{\displaystyle\sum\limits_{i=1}^{r}}
Y_{i}\otimes dc_{i}$ has the form%
\[%
{\displaystyle\sum\limits_{i=1}^{r}}
Y_{i}\otimes dc_{i}=\left[
\begin{array}
[c]{c|c}%
0_{2n} & 0_{2n\times r}\\\hline
0_{r\times2n} & I_{r}%
\end{array}
\right]
\]
and thus the partition of identity (\ref{partition}) is satisfied. Thus, all
the conditions of Definition \ref{dualdef} are satisfied. Notice also that
both $\Pi$ and $\Omega$ have constant rank $2n$. This simple example will be
further developed in Section 4.
\end{example}

We will now present a non-canonical example.

\begin{example}
\label{drugi}In our second example we consider a five dimensional manifold
$\mathcal{M}$ parametrized (locally) by coordinates $(q_{1},q_{2},p_{1}%
,p_{2},e)$ and a pair of operators given by%
\begin{equation}
\Pi=\left[
\begin{array}
[c]{ccccc}%
0 & 0 & 0 & \frac{1}{2}q_{1} & p_{1}\\
0 & 0 & \frac{1}{2}q_{1} & q_{2} & p_{2}\\
0 & -\frac{1}{2}q_{1} & 0 & -\frac{1}{2}p_{1} & 0\\
-\frac{1}{2}q_{1} & -q_{2} & \frac{1}{2}p_{1} & 0 & e\\
-p_{1} & -p_{2} & 0 & -e & 0
\end{array}
\right] \label{pi2}%
\end{equation}
and%
\begin{equation}
\Omega=\left[
\begin{array}
[c]{ccccc}%
0 & 2\frac{p_{1}}{q_{1}^{2}} & 4\frac{q_{2}}{q_{1}^{2}} & -2\frac{1}{q_{1}} &
\frac{1}{2}\left(  p_{1}p_{2}-q_{1}e\right) \\
-2\frac{p_{1}}{q_{1}^{2}} & 0 & -2\frac{1}{q_{1}} & 0 & -\frac{1}{2}p_{1}%
^{2}\\
-4\frac{q_{2}}{q_{1}^{2}} & 2\frac{1}{q_{1}} & 0 & 0 & \frac{1}{2}q_{1}%
p_{2}-q_{2}p_{1}\\
2\frac{1}{q_{1}} & 0 & 0 & 0 & \frac{1}{2}q_{1}p_{1}\\
\frac{1}{2}\left(  q_{1}e-p_{1}p_{2}\right)  & \frac{1}{2}p_{1}^{2} &
q_{2}p_{1}-\frac{1}{2}q_{1}p_{2} & -\frac{1}{2}q_{1}p_{1} & 0
\end{array}
\right] \label{omega2}%
\end{equation}
Both these operators have constant rank $4$ (generically). It is easy to check
that $\Pi$ is Poisson and that $\Omega$ is closed. The only (independent)
Casimir of $\Pi$ is given by%
\begin{equation}
c=-\frac{1}{2}q_{2}p_{1}^{2}+\frac{1}{2}q_{1}p_{1}p_{2}-\frac{1}{4}eq_{1}%
^{2},\label{casdrugi}%
\end{equation}
while the kernel of $\Omega$ is spanned by the vector field%
\[
Y=p_{1}\frac{\partial}{\partial q_{1}}+p_{2}\frac{\partial}{\partial q_{2}%
}+e\frac{\partial}{\partial p_{2}}-\frac{4}{q_{1}^{2}}\frac{\partial}{\partial
e}.
\]
Naturally, $Y(c)=1$. The explicit form of the tensor $Y\otimes dc$ is rather
complicated but a direct calculation shows that $\Pi\Omega+Y\otimes dc=I$.
Thus, $(\Pi,\Omega)$ is a dual P-p pair. This example will also be developed
later on.
\end{example}

The first example (Example \ref{pierwszy}) illustrates that the following
existence statement must be true.

\begin{proposition}
\label{existence}For a given Poisson operator $\Pi$ of constant rank on
$\mathcal{M}$ there always exists (locally) a closed two-form $\Omega$ such
that $(\Pi,\Omega)$ is a dual P-p pair and vice versa, for a given
presymplectic form $\Omega$ of constant rank there always exists a Poisson
operator $\Pi$ such that $(\Pi,\Omega)$ is a dual P-p pair.
\end{proposition}

\begin{proof}
Given a Poisson operator $\Pi$ (a closed two-form $\Omega$) it is enough to
pass to the Darboux coordinates for $\Pi$ ($\Omega$)and choose $\Omega$ ($\Pi
$) as in Example \ref{pierwszy}.
\end{proof}

\section{Dirac reduction of Poisson-presymplectic pairs}

Consider now a smooth $m$-dimensional manifold $\mathcal{M}$ endowed with a
dual P-p pair $(\Pi,\Omega)$ as in definitions \ref{dualdef} and \ref{Ppdef}
and a smooth $s$-dimensional foliation $\mathcal{S}$ of $\mathcal{M}$ with the
leaves $\mathcal{S}_{\nu}=\left\{  x\in\mathcal{M}:\varphi_{i}(x)=\nu
_{i}\text{, }\nu_{i}\in\mathbf{R}\text{, }i=1,\ldots,k\,\right\}  $ given by
$s$ functionally independent functions $\varphi_{i}:M\rightarrow\mathbf{R}$.
In this section we present a procedure of reducing of the dual P-p pair
$(\Pi,\Omega)$ to a dual P-p pair $(\pi_{R},\omega_{R})$ on any leaf
$\mathcal{S}_{\nu}$ of $\mathcal{S}$ provided that some additional assumptions
about relative positions of the foliations $\mathcal{N}$ and $\mathcal{S}$
hold and that we are in a generic case that will be called Dirac case. This
reduction will be similar to ideas developed by J. Marsden and T. Ratiu (see
\cite{MarsdenRatiu} and \cite{Diracrevisited}).

Let us thus fix a distribution $\mathcal{Z}$ (to be determined later) of
constant dimension $k=m-s$ (that is a smooth collection of $k$-dimensional
subspaces $\mathcal{Z}_{x}\subset$ $T_{x}\mathcal{M}$ at every point $x$ in
$\mathcal{M}$) that is transversal to $\mathcal{S}$ in the sense that no
vector field $Z\in\mathcal{Z}$ is at any point tangent to the foliation
$\mathcal{S}$. Hence we have
\[
T_{x}\mathcal{M}=T_{x}\mathcal{S}_{\nu}\oplus\mathcal{Z}_{x}%
\]
for every $x\in\mathcal{S}_{\nu}$, and similarly%
\[
\text{\ \ }T_{x}^{\ast}\mathcal{M}=T_{x}^{\ast}\mathcal{S}_{\nu}%
\oplus\mathcal{Z}_{x}^{\ast}%
\]
where $T_{x}^{\ast}\mathcal{S}$ is the annihilator of $\mathcal{Z}_{x}$ and
$\mathcal{Z}_{x}^{\ast}$ is the annihilator of $T_{x}\mathcal{S}$. This
distributions is assumed to be regular i.e. there exists linearly independent
vector fields $Z_{i}$, $i=1,\ldots,k$, such that $\mathcal{Z}=Sp\{Z_{i}%
\}_{i=1,\ldots k}$. Without loss of generality we can assume that the vector
fields $Z_{i}$ are chosen so that the following normalization condition holds%
\[
Z_{i}(\varphi_{j})=\delta_{ij}.
\]
There exists a natural projection $X_{||}$ of an arbitrary vector field $X$ on
$\mathcal{M}$ along $\mathcal{Z}$ onto the foliation $\mathcal{S}$ given by%
\[
X_{||}=X-%
{\textstyle\sum\limits_{i=1}^{k}}
X(\varphi_{i})Z_{i},
\]
as obviously $X_{||}(\varphi_{i})=0$. Similarly, any one-form $\alpha$ can be
naturally projected along $\mathcal{Z}$ to a one-form $a_{||}$ on $T^{\ast
}\mathcal{S}$ as follows:%
\begin{equation}
\alpha_{||}=\alpha-%
{\textstyle\sum\limits_{i=1}^{k}}
\alpha(Z_{i})d\varphi_{i}\label{projalfa}%
\end{equation}
since $\alpha_{||}(Z_{i})=0$. \ Finally, let us define the vector fields
$X_{i}$, $i=1,\ldots,s$ as the Hamiltonian vector fields%
\[
X_{i}=\Pi\,d\varphi_{i}.
\]

\begin{definition}
A function $F:\mathcal{M}\rightarrow\mathbf{R}$ is invariant with respect to
$\mathcal{Z}$ \ if $L_{Z}F=Z(F)=0$ for any $Z\in\mathcal{Z}$.
\end{definition}

\begin{definition}
A Poisson operator $\Pi$ is invariant with respect to the distribution
$\mathcal{Z}$ \ if $\left\{  F,G\right\}  _{\Pi}$ is a $\mathcal{Z}$-invariant
function for any pair of $\mathcal{Z}$-invariant functions $F$ and $G$. That
is, if $L_{Z_{i}}F=L_{Z_{i}}G=0$ then $L_{Z_{i}}\left\{  F,G\right\}  _{\Pi}=0
$.
\end{definition}

Let us now consider a $\mathcal{Z}$-invariant Poisson operator $\Pi$ and
define the following bilinear map:%
\begin{equation}
\Pi_{D}\left(  \alpha,\beta\right)  =\Pi(\alpha_{||},\beta_{||})\text{ \ for
any pair }\alpha,\beta\text{ of one-forms.}\label{defpi}%
\end{equation}

This new mapping induces a new bracket for functions on $\mathcal{M}$:%
\[
\left\{  F,G\right\}  _{\Pi_{D}}=\Pi_{D}\left(  dF,dG\right)  =\Pi(\left(
dF\right)  _{||},\left(  dG\right)  _{||})
\]
and thus it is easy to show that the corresponding bivector $\Pi_{D}$ has the
following form%
\begin{equation}
\Pi_{D}=\Pi-%
{\textstyle\sum\limits_{i}}
X_{i}\wedge Z_{i}+\frac{1}{2}%
{\textstyle\sum\limits_{i,j}}
\varphi_{ij}Z_{i}\wedge Z_{j}\label{pide}%
\end{equation}
which we can treat as a deformation of the original Poisson bivector $\Pi$.
Here the functions $\varphi_{ij}$ are defined as%
\begin{equation}
\varphi_{ij}=\left\{  \varphi_{i},\varphi_{j}\right\}  _{\Pi}=X_{j}%
(\varphi_{i}).\label{fij}%
\end{equation}

\begin{theorem}
For any $x\in\mathcal{M}$%
\[
\Pi_{D}(\alpha_{x})\in T_{x}\mathcal{S}\text{ \ \ for any }\alpha_{x}\in
T_{x}^{\ast}\mathcal{M}\text{, }%
\]
i.e. the image of $\Pi_{D\text{ }}$is tangent to the foliation $\mathcal{S}$.
\end{theorem}

\begin{proof}
We have to show that $\Pi_{D}(d\varphi_{k})=0$ for all $k$. According to
(\ref{pide}) we have%
\begin{align*}
\Pi_{D}(d\varphi_{k})  & =X_{k}-%
{\textstyle\sum\limits_{i}}
\delta_{ik}X_{i}+%
{\textstyle\sum\limits_{i}}
\varphi_{ki}Z_{i}+\frac{1}{2}%
{\textstyle\sum\limits_{i,j}}
\varphi_{ij}\left(  \delta_{jk}Z_{i}-\delta_{ik}Z_{j}\right)  =\\
& =%
{\textstyle\sum\limits_{i}}
\varphi_{ki}Z_{i}+\frac{1}{2}%
{\textstyle\sum\limits_{i}}
\varphi_{ik}Z_{i}-\frac{1}{2}%
{\textstyle\sum\limits_{j}}
\varphi_{kj}Z_{j}=0\text{,}%
\end{align*}
due to skewsymmetry of $\varphi_{ij}$.
\end{proof}

\begin{theorem}
If a Poisson operator $\Pi$ is $\mathcal{Z}$-invariant then the bivector
(\ref{pide}) is Poisson.
\end{theorem}

\begin{proof}
The operator $\Pi_{D}$ is obviously antisymmetric and the corresponding
bracket $\left\{  \cdot,\cdot\right\}  _{\Pi_{D}}$ satisfies Lebniz rule. The
Jacobi identity for $\left\{  \cdot,\cdot\right\}  _{\Pi_{D}}$%
\[
\left\{  \left\{  F,G\right\}  _{\Pi_{D}},H\right\}  _{\Pi_{D}}+\text{cycl.}=0
\]
reads due to (\ref{defpi}) as%
\begin{equation}
\left\langle \left(  d\left\langle \left(  dF\right)  _{||},\Pi\left(
dG\right)  _{||}\right\rangle \right)  _{||},\Pi\left(  dH\right)
_{||}\right\rangle +\text{cycl.}=0.\label{cycl}%
\end{equation}
But due to (\ref{projalfa})
\[
\left(  d\left\langle \left(  dF\right)  _{||},\Pi\left(  dG\right)
_{||}\right\rangle \right)  _{||}=d\left\langle \left(  dF\right)  _{||}%
,\Pi\left(  dG\right)  _{||}\right\rangle -%
{\textstyle\sum\limits_{i}}
Z_{i}\left(  \left\langle \left(  dF\right)  _{||},\Pi\left(  dG\right)
_{||}\right\rangle \right)  d\varphi_{i},
\]
and in every point $x\in\mathcal{M}$ we have $\left(  dF\right)  _{||}$ ,
$\left(  dG\right)  _{||}$ $\in$ $T^{\ast}\mathcal{S}$ so that
\[
Z_{i}\left(  \left\langle \left(  dF\right)  _{||},\Pi\left(  dG\right)
_{||}\right\rangle \right)  =\left\langle \left(  dF\right)  _{||},\left(
L_{Z_{i}}\Pi\right)  \left(  dG\right)  _{||}\right\rangle =0,
\]
the last equality is fulfilled due to the assumed $\mathcal{Z}$-invariance of
$\Pi$. Thus, the Jacobi identity (\ref{cycl}) reads actually as%
\[
\left\langle d\left\langle \left(  dF\right)  _{||},\Pi\left(  dG\right)
_{||}\right\rangle ,\Pi\left(  dH\right)  _{||}\right\rangle +\text{cycl.}=0,
\]
and is obviously satisfied due to the Jacobi identity for $\Pi$.
\end{proof}

Since the deformed operator $\Pi_{D}$ is Poisson and since the functions
$\varphi_{i}$ are its Casimirs, we can properly restrict $\Pi_{D}$ to any of
its symplectic leaves $\mathcal{S}_{\nu}$ obtaining a reduced Poisson operator
$\pi_{R\text{ }}$ on every leaf $\mathcal{S}_{\nu}.$%

\[
\pi_{R\text{ }}\overset{\text{def}}{=}\left.  \Pi_{D}\right\vert
_{\mathcal{S}_{\nu}}%
\]
Up to now the distribution $\mathcal{Z}$ was not fully determined. In the
generic case (that we call \emph{Dirac case}), when all the vector fields
$X_{i}$ are transversal to the foliation $\mathcal{S}$, we can choose the
distribution $\mathcal{Z}$ simply as the span of the vector fields $X_{i}$,
$\mathcal{Z}=Sp\{X_{i}\}_{i=1,..,k}$. We can now define our vector fields
$Z_{i}$ as a new basis of $\mathcal{Z}$:%
\begin{equation}
Z_{i}=%
{\textstyle\sum\limits_{j=1}^{k}}
(\varphi^{-1})_{ji}X_{j}\text{, \ \ }i=1,\ldots,k\label{Zdirac}%
\end{equation}
Indeed, since $\det(\varphi)\neq0$ the vector fields $Z_{i}$ also span the
distribution $\mathcal{Z}$ and moreover satisfy the normalization condition
$\left\langle d\varphi_{i},Z_{j}\right\rangle =Z_{j}(\varphi_{i})=\delta_{ij}%
$, as
\[
Z_{j}(\varphi_{i})=%
{\textstyle\sum\limits_{s=1}^{k}}
(\varphi^{-1})_{sj}X_{s}(\varphi_{i})=%
{\textstyle\sum\limits_{s=1}^{k}}
(\varphi^{-1})_{sj}\varphi_{is}=\delta_{ij}.
\]
Moreover, such choice of $Z_{i}$ makes the operator $\Pi$ $\mathcal{Z}%
$-invariant: if $L_{X_{i}}F=L_{X_{i}}G=0$ for all $i$ then $L_{X_{i}%
}\{F,G\}_{\Pi}=\left\langle dF,\left(  L_{X_{i}}\Pi\right)  \,dG\right\rangle
=0$ since $L_{X_{i}}\Pi=0$ ($\left\{  X_{i}\right\}  $ is just another basis
of $\mathcal{Z}$). In this case the deformation (\ref{pide}) attains the form:%
\begin{equation}
\Pi_{D}=\Pi-\frac{1}{2}\sum_{i=1}^{k}X_{i}\wedge Z_{i}\label{defdir}%
\end{equation}
and is, as mentioned above, Poisson. It is easy to check this operator defines
the following bracket on $C^{\infty}(\mathcal{M)}$
\begin{equation}
\{F,G\}_{\Pi_{D}}=\{F,G\}_{\Pi}-\sum_{i,j=1}^{k}\{F,\varphi_{i}\}_{\Pi
}(\varphi^{-1})_{ij}\{\varphi_{j},G\}_{\Pi},\label{Diracbracket}%
\end{equation}
(where $F,G:\mathcal{M}\rightarrow\mathbf{R}$ are two \emph{arbitrary}
functions on $\mathcal{M}$) which is just the well known \emph{Dirac
deformation} \cite{Dirac} of the bracket $\{.,.\}_{\Pi}$ associated with $\Pi
$. As we will show below in a more general context, $\ker(\Pi_{D}%
)=Sp\{d\varphi_{i},dc_{j}\}_{i=1,\ldots,k,\text{ }j=1,\ldots,r}$, i.e. the
Dirac deformation preserves all the old Casimir functions $c_{i}$\ and
introduces new Casimirs $\varphi_{i}$.

In the case when all the vector fields $X_{i}$ are tangent to the foliation
$\mathcal{S}$ (we call this case \emph{tangent case}) the foliation
$\mathcal{S}$ is Lagrangian with respect to any $\Omega$ dual to $\Pi$. Then
the deformation (\ref{pide}) attains the form%
\begin{equation}
\Pi_{D}=\Pi-%
{\textstyle\sum\limits_{i=1}^{k}}
X_{i}\wedge Z_{i},\label{deftan}%
\end{equation}
and has been considered in \cite{degiovanni} and in \cite{falquipedroni}.

Let us observe that the formula (\ref{pide}) can be rewritten as $\Pi_{D}%
=\Pi-\sum_{i}V_{i}\wedge Z_{i}$ with $V_{i}=X_{i}-\frac{1}{2}\sum_{j}%
\varphi_{ji}Z_{j}.$ A generalization of this formula has been considered in
\cite{Diracrevisited} where the vector fields $V_{i\text{ }}$ were determined
only up to a functional equation $V_{i}=X_{i}+\sum_{j}V_{j}(\varphi_{i})Z_{j}%
$. Two natural solutions to this equation in the above mentioned two cases are
$V_{i}=\frac{1}{2}X_{i}$ (in Dirac case) or $V_{i}=X_{i}$ (in tangent case)
and they yield exactly the two above deformations (\ref{defdir}) and
(\ref{deftan}) respectively.

In any case, our process of reducing the operator $\Pi$ to $\pi_{R}$ consists
of two steps: we first deform $\Pi$ to $\Pi_{D}$ and then reduce in a natural
way $\Pi_{D}$ to $\pi_{R}$ through a plain restriction: $\pi_{R}=\Pi
_{D}|_{\mathcal{S}}$.

Our construction generalizes the construction of Marsden and Ratiu in the
following sense. Marsden and Ratiu presented in \cite{MarsdenRatiu} a natural
way of reducing of a given Poisson bracket $\left\{  \cdot,\cdot\right\}
_{\Pi}$ on $\mathcal{M}$ \ to a Poisson bracket $\left\{  \cdot,\cdot\right\}
_{\pi_{R}}$ on a given submanifold $\mathcal{S}_{0}$ (in our notation). Their
method is non-constructive in the sense that in order to find the bracket
$\left\{  f,g\right\}  _{\pi_{R}}$ of two functions $f$,$g:S_{0}\rightarrow R$
one has to calculate $\mathcal{Z}|_{\mathcal{S}_{0}}$-invariant prolongations
of these functions. Our construction is performed on the level of bivectors
rather than on the level of Poisson brackets. This construction (by
deformation of the bivector $\Pi$) applies directly to every leaf of the
distribution $\mathcal{S}$ \ and moreover it is constructive. At every leaf,
however, both constructions are equivalent, as it is easy to see. Also, our
construction can be extended to a similar construction for closed two-forms,
as it is shown below. On the other hand, we make the assumption about the
transversality of the distribution $\mathcal{Z}$ \ that was not present in the
original paper of Marsden and Ratiu. This assumption is however very natural
since it makes all the assumption of Poisson Reduction Theorem in
\cite{MarsdenRatiu} automatically satisfied.

Now we turn to an analogous question of reducing closed two-forms onto the
foliation $\mathcal{S}$. Of course, there always exists a natural restriction
of any closed two-form on any submanifold $\mathcal{S}_{\nu}$, obtained simply
by restricting its domain to $T\mathcal{S}_{\nu}$. However, in the case that
our closed two-form is a part of a dual P-p pair $(\Pi,\Omega)$ it is also
natural to consider a similar two-step procedure, where we first deform
$\Omega$ to $\Omega_{D}$ (such that $(\Pi_{D},\Omega_{D})$ is again a dual
pair) and then restrict $\Omega_{D}$ to a closed two-form $\omega_{R}$ on
$\mathcal{S}_{\nu}$ such that $(\pi_{R},\omega_{R})$ is a dual P-p pair. This
is the main aim of this paper.

Let us thus define, in analogy with (\ref{defpi}), the following bilinear map:%
\begin{equation}
\Omega_{D}(U,V)=\Omega(U_{||},V_{||})\text{ for any vector fields }U,V\text{
on }\mathcal{M}\text{.}\label{defomega}%
\end{equation}
This map induces the following two-form $\Omega_{D}$ on $\mathcal{M}$:%
\begin{equation}
\Omega_{D}=\Omega-%
{\textstyle\sum\limits_{i=1}^{k}}
\xi_{i}\wedge d\varphi_{i}-\frac{1}{2}%
{\textstyle\sum\limits_{i,j=1}^{k}}
\xi_{i}(Z_{j})d\varphi_{i}\wedge d\varphi_{j}\label{omegade}%
\end{equation}
where the one-forms $\xi_{i}$ are defined as%
\[
\xi_{i}=\Omega(Z_{i})\text{.}%
\]

The two-form $\Omega_{D}$ obviously restricts to the same two-form on
$\mathcal{S}_{\nu}$ as $\Omega$ does. That is, we can define a form
$\omega_{R}$ on every leaf of $\mathcal{S}$ through the plain restriction of
$\Omega_{D}$ (or $\Omega$) to $\mathcal{S}_{\nu}$:%
\[
\omega_{R}\overset{\text{def}}{=}\Omega_{D}|_{\mathcal{S}_{\nu}}\equiv
\Omega|_{\mathcal{S}_{\nu}}.
\]
It is obvious that $\Omega_{D}|_{\mathcal{S}_{\nu}}\equiv\Omega|_{\mathcal{S}%
_{\nu}}$ since the last two terms in (\ref{omegade}) vanish on $\mathcal{S}%
_{\nu}$.

Let us now assume that $(\Pi,\Omega)$ is a dual P-p pair in the sense of
definitions \ref{dualdef} and \ref{Ppdef}. We will show that in the generic
(Dirac) case and under certain conditions both pairs $(\Pi_{D},\Omega_{D})~
$\ and $(\pi_{R},\omega_{R})$ are dual pairs.

\begin{theorem}
\label{obrot}Suppose that $(\Pi,\Omega)$ is a dual P-p pair with $\ker
(\Pi)=Sp\{dc_{i}\}_{i=1,..,r}$, $\ker(\Omega)=Sp\{Y_{i}\}_{i=1,..,r}$ and with
the corresponding foliation $\mathcal{N}$ of $\mathcal{M}$. Suppose also that
the constraints $\varphi_{j}$, $j=1,\ldots,k$ define a foliation $\mathcal{S}$
of $\mathcal{M}$ with some transversal distribution spanned by the vector
fields $Z_{i}$ such that $\Pi$ is $\mathcal{Z}$-invariant and such that
$Z_{i}(\varphi_{j})=\delta_{ij}$. Suppose also that all $Y_{i}$ are tangent to
$\mathcal{S}$ (i.e. $Y_{i}(\varphi_{j})=0$ for all $i,j$) and that all $Z_{i}$
are tangent to $\mathcal{N}$ \ (i.e. $Z_{i}(c_{j})=0$ for all $i,j$). Then the
pair $(\Pi_{D},\Omega_{D})$ given by (\ref{pide}) and (\ref{omegade}) has the
following properties\newline

\begin{enumerate}
\item $\ker(\Pi_{D})=Sp\{dc_{i},d\varphi_{j}\}$, $\ \ \ker(\Omega
_{D})=Sp\{Y_{i},Z_{j}\}$

\item In the generic (Dirac) case when $Z_{i}$ are obtained as in
(\ref{Zdirac}) the pair $(\Pi_{D},\Omega_{D})$ is a dual pair.
\end{enumerate}
\end{theorem}

\begin{proof}
We have already showed that $\Pi_{D}\,d\varphi_{i}=0$. A similar computation
yields that $\Pi_{D}\,dc_{i}=0$. Using (\ref{omegade}) we obtain%
\begin{align*}
\Omega_{D}(Y_{j})  & =\Omega(Y_{j})-%
{\textstyle\sum_{i=1}^{k}}
Y_{j}(\varphi_{i})\xi_{i}+%
{\textstyle\sum_{i=1}^{k}}
\left\langle \xi_{i},Y_{j}\right\rangle d\varphi_{i}-\\
& -\frac{1}{2}%
{\textstyle\sum_{i,l=1}^{k}}
\xi_{i}(Z_{l})\left(  Y_{j}(\varphi_{l})d\varphi_{i}-Y_{j}(\varphi
_{i})d\varphi_{i}\right) \\
& =0,
\end{align*}
since $\Omega(Y_{j})=0$, $\left\langle \xi_{i},Y_{j}\right\rangle
=\Omega(Z_{i},Y_{j})=-\Omega(Y_{j},Z_{i})=0$ and $Y_{j}(\varphi_{i})=0$ by
assumption. Further%
\[
\Omega_{D}(Z_{j})=\xi_{j}-%
{\textstyle\sum_{i=1}^{r}}
Z_{j}(\varphi_{i})\xi_{i}+%
{\textstyle\sum_{i=1}^{k}}
\left\langle \xi_{i},Z_{j}\right\rangle d\varphi_{i}-\frac{1}{2}%
{\textstyle\sum_{i,l=1}^{k}}
\left\langle \xi_{i},Z_{l}\right\rangle \left(  \delta_{jl}d\varphi_{i}%
-\delta_{ji}d\varphi_{l}\right)  =0
\]
all due to $Z_{j}(\varphi_{i})=\delta_{ji}$. This concludes the proof of the
first statement. Using some elementary tensor relations one can (after some
direct but cumbersome calculations) show that the pair $(\Pi_{D},\Omega_{D})$
satisfies the following identity on $T\mathcal{M}$:%
\begin{equation}
I=\Pi_{D}\,\Omega_{D}+%
{\textstyle\sum\limits_{i=1}^{r}}
Y_{i}\otimes dc_{i}+%
{\textstyle\sum\limits_{j=1}^{k}}
Z_{j}\otimes d\varphi_{j}-T\label{newpart}%
\end{equation}
where the (1,1)-tensor $T$ is of the form%
\[
T=%
{\textstyle\sum\limits_{i=1}^{k}}
\left(  X_{i}+%
{\textstyle\sum\limits_{j=1}^{k}}
\varphi_{ij}Z_{i}\right)  \otimes\left(  \xi_{i}+%
{\textstyle\sum_{l=1}^{k}}
\xi_{l}\left(  Z_{i}\right)  d\varphi_{l}\right)  .
\]
In the Dirac case the expresssions in both parentheses in $T$ vanish for every
$i$ due to (\ref{Zdirac}) so that the whole tensor $T$ vanishes. Further, the
relations $Y_{i}(\varphi_{j})=0$, $Y_{i}(c_{j})=\delta_{ij}$, $Z_{i}%
(\varphi_{j})=\delta_{ij}$, $Z_{i}(c_{j})=0$ are just part of our assumptions.
Thus, in the Dirac case all the requirements of Definition \ref{dualdef} are satisfied.
\end{proof}

We are now in position to prove the main theorem of this paper.

\begin{theorem}
The pair $(\pi_{R},\omega_{R})$ obtained through the restriction $(\pi
_{R},\omega_{R})=(\Pi_{D},\Omega_{D})|_{\mathcal{S}_{\nu}}$ of $(\Pi
_{D},\Omega_{D})$ given by (\ref{pide}) and (\ref{omegade}) is in the Dirac
case a dual P-p pair on every leaf $\mathcal{S}_{\nu}$ of $\mathcal{S}$ with
the Casimirs $c_{i}|_{\mathcal{S}_{\nu}}$ , the kernel of $\omega_{R}$ spanned
by $Y_{i}$ (notice that $Y_{i}$ are tangent to $\mathcal{S}_{\nu}$) and with
the partition of identity on $T\mathcal{S}_{\nu}$ given by%
\begin{equation}
I=\pi_{R}\omega_{R}+%
{\displaystyle\sum\limits_{i=1}^{r}}
Y_{i}\otimes d(c_{i}|_{\mathcal{S}_{\nu}}).\label{partonS}%
\end{equation}

\end{theorem}

\begin{proof}
The proof of this theorem follows from the fact that $(\pi_{R},\omega
_{R})=(\Pi_{D},\Omega_{D})|_{\mathcal{S}_{\nu}}$. The partition of identity
(\ref{partonS}) follows from the partition (\ref{newpart}) since $d\varphi
_{i}|_{\mathcal{S}_{\nu}}=0$ and since $T=0$ in the Dirac case. Further,
$\pi_{R}$ is Poisson since $\Pi_{D\,}$ is. We only have to check that
$\omega_{R}$ is closed.

Obviously, $\Omega_{D}$ is usually not closed, as according to (\ref{omegade})
in the Dirac case we easily obtain
\[
d\Omega_{D}=\frac{1}{2}%
{\textstyle\sum_{i=1}^{k}}
d\varphi_{i}\wedge d\xi_{i}\neq0\text{.}%
\]
However, $\omega_{R}=\Omega_{D}|_{\mathcal{S}_{\nu}}$ so that $d\omega
_{R}=d(\Omega_{D}|_{\mathcal{S}_{\nu}})=(d\Omega_{D})|_{\mathcal{S}_{\nu}}=0$
due to the above formula, again since $d\varphi_{i}|_{\mathcal{S}_{\nu}}=0$.
\end{proof}

Thus, starting from a dual P-p pair $(\Pi,\Omega)$ and a proper foliation
$\mathcal{S}$ (defined by a Dirac second-class constraints $\varphi_{i}$) we
have constructed a dual P-p pair $(\pi_{R},\omega_{R})$ on every leaf of
$\mathcal{S}$.

\section{Examples}

Let us now continue with Example \ref{pierwszy} and Example \ref{drugi}. To
illustrate our approach we will also construct the deformations $\Omega_{D}$
dual to the respective bivectors $\Pi_{D}$, although it is not necessary for
the actual construction of the reductions $\omega_{R}$ that can be obtained
directly by restricting $\Omega$ to $\mathcal{S}_{\nu}$.

\begin{example}
\label{pierwszykont}(Ex \ref{pierwszy} continued). Assume that $n=3$ and $r=1
$ so that the manifold $\mathcal{M}$ is of dimension 7 and the local
coordinates are $(q_{1},q_{2},q_{3},p_{1},p_{2},p_{3},c)$. The original dual
P-p pair $(\Pi,\Omega)$ is given by (\ref{piomega1}). Let us now introduce a
5-dimensional submanifold $\mathcal{S}_{0}$ through the following pair of
constraints:%
\begin{equation}
\varphi_{1}\equiv q_{1}q_{2}+q_{3}=0\text{, \ }\varphi_{2}=p_{1}+p_{2}%
q_{1}+p_{3}q_{2}=0\label{wiezy1}%
\end{equation}
(for some motivation on the source of these constraints, see
(\cite{Diracpencils})) so that $k=2$ here. The constraints (\ref{wiezy1}) do
not contain the Casimir function $c$ explicitly so that the condition
$Y(\varphi_{i})=0$ is satisfied as $Y=\frac{\partial}{\partial c}$. The
$2\times2$ matrix $\varphi$ has the form:%
\[
\varphi=(2q_{2}+q_{1}^{2})\left[
\begin{array}
[c]{cc}%
0 & 1\\
-1 & 0
\end{array}
\right]
\]
The vector fields $X_{i}$ and then $Z_{i}$ can be easily computed
(\ref{Zdirac}). The result is%
\[
X_{1}=-q_{2}\frac{\partial}{\partial p_{1}}-q_{1}\frac{\partial}{\partial
p_{2}}-\frac{\partial}{\partial p_{3}}\text{, \ }X_{2}=\frac{\partial
}{\partial q_{1}}+q_{1}\frac{\partial}{\partial q_{2}}+q_{2}\frac{\partial
}{\partial q3}-p_{2}\frac{\partial}{\partial p_{1}}-p_{3}\frac{\partial
}{\partial p_{2}}%
\]%
\[
Z_{1}=\frac{1}{\varphi_{12}}X_{2}\text{, \ }Z_{2}=-\frac{1}{\varphi_{12}}%
X_{1},
\]
and one can see that $Z_{i}(c)=0$ as $Z_{i}$ do not contain derivation with
respect to the coordinate variable $c$. A direct computation of the expression
(\ref{pide}) leads to%
\[
\Pi_{D}=\frac{1}{\varphi_{12}}\left[
\begin{array}
[c]{c|c}%
\begin{array}
[c]{cc}%
0_{3} & A\\
-A^{t} & B
\end{array}
& 0_{6\times1}\\\hline
0_{1\times6} & 0
\end{array}
\right]  ,
\]
with
\[
A=\left[
\begin{array}
[c]{ccc}%
q_{2}+q_{1}^{2} & -q_{1} & -1\\
-q_{1}q_{2} & 2q_{2} & -q_{1}\\
-q_{2}^{2} & -q_{1}q_{2} & q_{2}+q_{1}^{2}%
\end{array}
\right]  \text{, \ \ }B=\left[
\begin{array}
[c]{ccc}%
0 & p_{2}q_{1}-p_{3}q_{2} & p_{2}\\
p_{3}q_{2}-p_{2}q_{1} & 0 & p_{3}\\
-p_{2} & -p_{3} & 0
\end{array}
\right]  .
\]
It can be easily shown that $\Pi_{D\text{ }}$ is indeed Poisson. The one-forms
$\xi_{i}$ are given by%
\[
\xi_{1}=\frac{1}{\varphi_{12}}d\varphi_{2}\text{, }\xi_{2}=-\frac{1}%
{\varphi_{12}}d\varphi_{1}%
\]
and (\ref{omegade}) yields%
\[
\Omega_{D}=\frac{1}{\varphi_{12}}\left[
\begin{array}
[c]{c|c}%
\begin{array}
[c]{cc}%
-B & A^{t}\\
-A & 0
\end{array}
& 0_{6\times1}\\\hline
0_{1\times6} & 0
\end{array}
\right]  ,
\]
One can easily check that $d\Omega_{D}\neq0$. However, $(\Pi_{D},\Omega_{D})$
is a dual pair with $Y=\frac{\partial}{\partial c}$. In order to obtain
explicit expressions on $\pi_{R}$ and $\omega_{R}$ we pass to new coordinate
system $(q_{1},q_{2},\varphi_{1},\varphi_{2},p_{2},p_{3},c)$ as the
constraints (\ref{wiezy1}) are in a natural way soluble with respect to
$q_{3},p_{1}$. In these Casimir variables our operators attain the form%
\[
\Pi_{D}^{\prime}=\frac{1}{\varphi_{12}}\left[
\begin{array}
[c]{c|c}%
\begin{array}
[c]{cc}%
0_{3} & A^{\prime}\\
-A^{\prime t} & B^{\prime}%
\end{array}
& 0_{6\times1}\\\hline
0_{1\times6} & 0
\end{array}
\right]  ,
\]
with
\[
A^{\prime}=\left[
\begin{array}
[c]{ccc}%
0 & -q_{1} & -1\\
0 & 2q_{2} & -q_{1}\\
0 & 0 & 0
\end{array}
\right]  \text{, \ \ }B^{\prime}=\left[
\begin{array}
[c]{ccc}%
0 & 0 & 0\\
0 & 0 & p_{3}\\
0 & -p_{3} & 0
\end{array}
\right]  .
\]
and%
\[
\Omega_{D}^{\prime}=\left[
\begin{array}
[c]{c|c}%
\begin{array}
[c]{cc}%
C^{\prime} & D^{\prime}\\
-D^{\prime t} & 0_{3}%
\end{array}
& 0_{6\times1}\\\hline
0_{1\times6} & 0
\end{array}
\right]  ,
\]
with%
\[
C^{\prime}=\left[
\begin{array}
[c]{ccc}%
0 & p_{3} & 0\\
-p_{3} & 0 & 0\\
0 & 0 & 0
\end{array}
\right]  \text{, \ \ }D^{\prime}=\left[
\begin{array}
[c]{ccc}%
-1 & q_{1} & 2q_{2}\\
0 & -1 & q_{1}\\
\frac{1}{\varphi_{12}} & 0 & -1
\end{array}
\right]  .
\]
Now, if we parametrize the submanifold $\mathcal{S}_{0}$ with the coordinates
$(q_{1},q_{2},p_{2},p_{3},c)$ then we can immediately obtain the expressions
for $\pi_{R}$ and $\omega_{R}$:
\[
\pi_{R}=\frac{1}{\varphi_{12}}\left[
\begin{array}
[c]{ccccc}%
0 & 0 & -q_{1} & -1 & 0\\
0 & 0 & 2q_{2} & -q_{1} & 0\\
q_{1} & -2q_{2} & 0 & p_{3} & 0\\
1 & q_{1} & -p_{3} & 0 & 0\\
0 & 0 & 0 & 0 & 0
\end{array}
\right]  \text{ , }\omega_{R}=\left[
\begin{array}
[c]{ccccc}%
0 & p_{3} & q_{1} & 2q_{2} & 0\\
-p_{3} & 0 & -1 & q_{1} & 0\\
-q_{1} & -1 & 0 & 0 & 0\\
-2q2 & q_{1} & 0 & 0 & 0\\
0 & 0 & 0 & 0 & 0
\end{array}
\right]
\]
and it can be checked directly that it is indeed a dual P-p pair. For example,
one can immediately check that%
\[
I=\pi_{R}\omega_{R}+Y\otimes d(c|_{\mathcal{S}_{0}}).
\]

\end{example}

The presented example is very simple, but illustrative. Let us now turn to our
non-canonical example.

\begin{example}
\label{drugikont}(Ex \ref{drugi} continued) This time we work with the dual
P-p pair (\ref{pi2})-(\ref{omega2}) written in coordinates $(q_{1},q_{2}%
,p_{1},p_{2},e)$. Observe that now $e$ is not any Casimir variable (and hence
a different letter to denote this odd variable). Let us now introduce a
3-dimensional submanifold $\mathcal{S}_{0}$ through the following pair of
constraints:%
\begin{equation}
\varphi_{1}\equiv p_{1}-1=0\text{, \ }\varphi_{2}\equiv-p_{2}p_{1}^{2}%
+eq_{1}p_{1}+2\ln(q_{1}^{2})=0\label{wiezy2}%
\end{equation}
so that $k=2$ again. It is easy to check that $Y(\varphi_{i})=0$ and only a
bit more difficult to see that $Z_{i}(c)=0$ so that the assumptions of Theorem
\ref{obrot} are again satisfied. The calculations similar to those in Example
\ref{pierwszykont} lead to%
\[
\Pi_{D}=\left[
\begin{array}
[c]{ccccc}%
0 & \frac{q_{1}^{2}}{2p_{1}} & 0 & q_{1} & p_{1}\\
-\frac{q_{1}^{2}}{2p_{1}} & 0 & 0 & A & p_{2}+\frac{4}{p_{1}^{2}}\\
0 & 0 & 0 & 0 & 0\\
-q_{1} & -A & 0 & 0 & e+\frac{4}{q_{1}p_{1}}\\
-p_{1} & -p_{2}-\frac{4}{p_{1}^{2}} & 0 & -e-\frac{4}{q_{1}p_{1}} & 0
\end{array}
\right]  ,
\]
($p_{1}$ is now a Casimir for $\Pi_{D}$) with $A=\frac{q_{1}p_{2}}{p_{1}%
}+\frac{2q_{1}}{p_{1}^{3}}-\frac{eq_{1}^{2}}{2p_{1}^{2}}$ and to%
\[
\Omega_{D}=\left[
\begin{array}
[c]{ccccc}%
0 & 2\frac{p_{1}}{q_{1}^{2}} & B & -2\frac{1}{q_{1}} & \frac{1}{2}p_{1}%
p_{2}-\frac{1}{2}eq_{1}\\
-2\frac{p_{1}}{q_{1}^{2}} & 0 & -2\frac{1}{q_{1}} & 0 & -\frac{1}{2}p_{1}%
^{2}\\
-B & 2\frac{1}{q_{1}} & 0 & 0 & C\\
2\frac{1}{q_{1}} & 0 & 0 & 0 & \frac{1}{2}q_{1}p_{1}\\
\frac{1}{2}eq_{1}-\frac{1}{2}p_{1}p_{2} & \frac{1}{2}p_{1}^{2} & -C &
-\frac{1}{2}q_{1}p_{1} & 0
\end{array}
\right]  ,
\]
with $B=4\frac{q_{2}}{q_{1}^{2}}-2\frac{e}{p_{1}^{2}}-8\frac{q_{1}}{p_{1}^{3}%
}$, $C=p_{1}q_{2}-\frac{1}{2}q_{1}p_{2}-2\frac{q_{1}}{p_{1}^{2}}$. It turns
out that in this case $\Omega_{D}$ is closed so that here our pair $(\Pi
_{D},\Omega_{D})$ is a dual P-p pair. In order to reduce this pair onto
$\mathcal{S}$ \ we pass to the Casimir variables $(q_{1},q_{2},\varphi
_{1},\varphi_{2},c)$ defined by constraints (\ref{wiezy2}) together with
(\ref{casdrugi})
\[
\left\{
\begin{array}
[c]{l}%
\varphi_{1}=p_{1}-1\\
\varphi_{2}=-p_{2}p_{1}^{2}+eq_{1}p_{1}+2\ln(q_{1}^{2})\\
c=-\frac{1}{2}q_{2}p_{1}^{2}+\frac{1}{2}q_{1}p_{1}p_{2}-\frac{1}{4}eq_{1}^{2}%
\end{array}
\right.
\]
It is possible to solve these equations with respect to $p_{1},p_{2},e$. We
get the expressions%
\begin{equation}
p_{1}=p_{1}(q,\varphi,c)=\varphi_{1}+1\text{, }p_{2}=p_{2}(q,\varphi,c)\text{,
}e=e(q,\varphi,c)\label{odwrotne}%
\end{equation}
that are however too complicated to present them explicitly. In these new
variables the operator $\Pi_{D}$ attains an almost canonical form%
\[
\Pi_{D}^{\prime}=\frac{q_{1}^{2}}{2(\varphi_{1}+1)}\left[
\begin{array}
[c]{c|c}%
\begin{array}
[c]{cc}%
0 & 1\\
-1 & 0
\end{array}
& 0_{2\times3}\\\hline
0_{3\times2} & 0_{3\times3}%
\end{array}
\right]
\]
while $\Omega_{D}$ attains a rather complicated form.
\[
\Omega_{D}^{\prime}=\left[
\begin{array}
[c]{ccccc}%
0 & -2\frac{p_{1}}{q_{1}^{2}} & 2\frac{e}{p_{1}^{2}}-4\frac{q_{2}}{q_{1}^{2}}
& -\frac{2}{p_{1}^{2}q_{1}} & \frac{2p_{1}p_{2}}{q_{1}^{2}}\\
2\frac{p_{1}}{q_{1}^{2}} & 0 & -\frac{2}{q_{1}} & 0 & -2\frac{p_{1}^{2}}%
{q_{1}^{2}}\\
4\frac{q_{2}}{q_{1}^{2}}-2\frac{e}{p_{1}^{2}} & \frac{2}{q_{1}} & 0 & \frac
{2}{p_{1}^{3}} & A^{\prime}\\
\frac{2}{p_{1}^{2}q_{1}} & 0 & -\frac{2}{p_{1}^{3}} & 0 & -\frac{2}{q_{1}%
p_{1}}\\
-\frac{2p_{1}p_{2}}{q_{1}^{2}} & 2\frac{p_{1}^{2}}{q_{1}^{2}} & -A^{\prime} &
\frac{2}{q_{1}p_{1}} & 0
\end{array}
\right]  ,
\]
with $A^{\prime}=-2\frac{p_{2}}{q_{1}}+2\frac{e}{p_{1}}-4\frac{p_{1}q_{2}%
}{q_{1}^{2}}$ and with $p_{1},$ $p_{2},e$ given by (\ref{odwrotne}). We are
now ready to reduce our operators onto $\mathcal{S}_{0}$. The result is
\[
\pi_{R}=\frac{q_{1}^{2}}{2}\left[
\begin{array}
[c]{c|c}%
\begin{array}
[c]{cc}%
0 & 1\\
-1 & 0
\end{array}
& 0_{2\times1}\\\hline
0_{1\times2} & 0
\end{array}
\right]  \text{ , }\omega_{R}=\frac{2}{q_{1}^{2}}\left[
\begin{array}
[c]{ccc}%
0 & -1 & B\\
1 & 0 & -1\\
B & 1 & 0
\end{array}
\right]
\]
with $B=(4c+2q_{2}-2q_{1}\ln(q_{1}^{2}))/q_{1}$. Again it can be checked
directly that it is indeed a dual P-p pair.
\end{example}

\section{Conclusions}

In this paper we have constructed a theory of Dirac-type reductions for
Poisson bivectors and presymplectic (i.e. closed but in general degenerate)
two-forms by embedding them in a geometrical object that we call 'dual pair'.
We systematically constructed the theory of dual pairs and of their special
type: Poisson-presymplectic pairs (i.e. dual pairs consisting of one Poisson
operator and one closed two-form). Using this theory we presented how to
project in principle any dual P-p pair onto submanifolds in such a way that
the reduced pair is again a dual P-p pair. Our method is in a sense a
generalization of \ the concepts of P.A.M. Dirac , J. Marsden and T. Ratiu. We
concluded the article by two examples: one starting from a canonical dual pair
and one non-canonical.

\bigskip

\end{document}